\documentclass[12pt]{article}
\usepackage{amssymb}
\usepackage{graphicx}

\begin{document}

\begin{titlepage}
\vskip 3cm
\begin{center}
Lithuanian Journal of Physics, 2001, \textbf{41}, No. 4-6,
551-555.
\vskip 2cm

\textbf{Modelling share volume traded in financial markets}

\vskip 1cm

\textbf{V. Gontis}

\vskip 1cm

\textit{Institute of Theoretical Physics and Astronomy,}

\textit{A. Go\v{s}tauto 12, 2600 Vilnius, Lithuania}

\textit{E-mail: gontis@ktl.mii.lt}

\vskip 1cm

\begin{quotation}
\textbf{Abstract.} \ A simple analytically solvable model exhibiting a $1/f$
\ spectrum in an arbitrarily wide frequency range was recently proposed by
Kaulakys and Me\v{s}kauskas (KM). Signals consisting of a sequence of pulses
show that inherent origin of the $1/f$ \ noise is Brownian fluctuations of
the average intervent time between subsequent pulses of the pulse sequence.
We generalize the KM model to reproduce the variety of self-affine time
series exhibiting power spectral density $S(f)$ scaled as power of their
frequency $f$. Numerical calculations\ with the generalized discrete model
(GDM) reproduce power spectral density $S(f)$ scaled as power of frequency $%
1/f^{\beta }$ for various values of $\beta $, including $\beta =1/2$ for
applications in financial markets. The particular applications of the model
proposed are related with financial time series of share volume traded.
\end{quotation}

\vfill
\end{center}
\end{titlepage}

\section{Introduction}

Physicists have recently begun doing research in finance with wide
application of models earlier introduce in physics \cite{1,2} . The
statistical properties of financial time series are attracting experts of
statistical physics, fluctuations analysis, dynamical chaos and others.
Papers on finance are appearing with some frequency in physics journals. A
new movement called \textit{econophysics} has been established \cite{2}. On
the other hand, several empirical studies have determined scale-invariant
behavior\ of both the long range correlations of price volatility and share
volume traded in financial markets \cite{3,4,5,6}. Mandelbrot introduced the
concept of fractals in terms of statistical self similarity \cite{7}\ and
using the context of self-affine time series extended the concept to time
series \cite{8}. The basic definition of self-affine time series is that the
power spectral density of the time series has a power-law dependence on
frequency $S(f)=f^{-\beta }$. Universality of $1/f$ noise, when $\beta =1$,
has led to speculations that there might exist some generic mechanism
underlying production of so general statistical properties. Such concepts as
fractional Brownian motion provide a formal procedure how to produce
self-affine time series, but can't serve as generic mechanism. We will
generalize the simple model introduced by Kaulakys and Me\v{s}kauskas (KM)
\cite{9}\ to generate time series in the range $0\leq \beta \leq 2$ with
particular interest in financial time series.

Long range correlations in time series $I(t)$ are quantified by
autocovariance (autocorrelation) function $C(s)$:
\begin{equation}
C(s)=C(-s)=\left\langle \frac{1}{T}\int_{0}^{T-s}I(t)I(t+s)dt\right\rangle ,
\end{equation}
with Wiwner-Khintchine relation to power spectral density $S(f)$ defined as:
\begin{equation}
S(f)=\lim_{T\rightarrow \infty }\left\langle \frac{2}{T}\left|
\int_{0}^{T}I(t)e^{-i2\pi ft}dt\right| ^{2}\right\rangle
=4\lim_{T\rightarrow \infty }\int_{0}^{T}C(s)\cos (2\pi fs)ds,
\end{equation}
where $T$ denotes the considered time interval of time series $I(t)$ . The
KM model is based on the time series generated as:
\begin{equation}
I(t)=\sum_{k}q_{k}\delta (t-t_{k})
\end{equation}
where $q_{k}$ is a contribution to the signal of one pulse and noise is due
to the correlations between the occurrence times $t_{k}$ of $\delta $ type
pulses. This model corresponds to the flow of point objects: photons,
electrons, cars, trades in financial markets and so on. The simplest version
of the model consist of sequence of transit times $t_{k}$ described by
recurrence equations:
\begin{eqnarray}
t_{k} &=&t_{k-1}+\tau _{k},  \nonumber \\
\tau _{k} &=&\tau _{k-1}-\gamma (\tau _{k-1}-\overline{\tau })+\sigma
\varepsilon _{k}.
\end{eqnarray}
Here\ the recurrence time $\tau _{k}=t_{k}-t_{k-1}$ fluctuates due to the
external random perturbation of the system by sequence of uncorrelated
normally distributed random variables $\left\{ \varepsilon _{k}\right\} $
with zero expectation and unit variance, where $\sigma $ denotes the
standard deviation of the white noise, $\gamma \ll 1$ is the recurrence time
$\tau $ relacsation rate to the some average value $\overline{\tau }$. Note
that $\tau $ follows an autoregressive process AR(1).

This model containing only one relaaxation time $\gamma ^{-1}$ can for
sufficiently small parameters $\gamma $ produce an exact $1/f$ - like
spectrum in wide range of frequence \cite{9}:
\begin{equation}
S(f)=\frac{\alpha _{H}}{\overline{\overline{\tau }}^{2}f},\quad f_{1}<f<\min
(f_{2},f_{\overline{\tau }}),
\end{equation}

where $\alpha _{H}$ is a dimensionless constant --- the Hooge parameter:

\begin{equation}
\alpha _{H}=\frac{2}{\sqrt{\pi }}Ke^{-K^{2}},\qquad K=\frac{\overline{\tau }%
\sqrt{\gamma }}{\sigma }.
\end{equation}
and $f_{1}=\gamma ^{3/2}/\pi \sigma $, $f_{2}=2\gamma ^{1/2}/\pi \sigma $, $%
f_{\tau }=(2\pi \overline{\tau })^{-1}.$

Here we present generalized KM model with particular interest to reproduce
the statistical properties of share volume traded in financial markets. We
will generate discrete share volume time series with recurrence time of
distinct trades described as AR(2) process. This particular application of
the model proposed by B. Kaulakys and T. Me\v{s}kauskas will exhibit
univesality of the mechanism underlying production of an $f^{-\beta }$ noise.

\section{Model definition}

In the numerical data analysis we usually deal with discrete sets of data.
There is the first need to modify the model to the discrete one. Let us
introduce a new conventional time scale defined with a time interval $\tau
_{d}.$ Integrating continuous signal $I(t)$ in the subsequent intervals of
length $\tau _{d}$ we will get discrete time series (let us call it volume $%
V_{r}$) :
\begin{equation}
V_{r}=\int_{t_{r}}^{t_{r}+\tau }I(t)dt={\sum_{k}}q_{k},\quad t_{r}=r\tau
_{d}.
\end{equation}
Consequently, the discrete power spectral density can be expressed as:
\begin{equation}
S(f_{s})=\left\langle \frac{2}{\tau _{d}n}\left|
\sum_{r=1}^{n}V_{r}e^{-i2\pi (s-1)(r-1)/n}\right| ^{2}\right\rangle ,
\end{equation}
where $f_{s}=\frac{s-1}{T}$, $T=\tau _{d}n$.

\begin{figure}[tbp]
  \begin{center}
  \includegraphics [width=0.8\hsize] {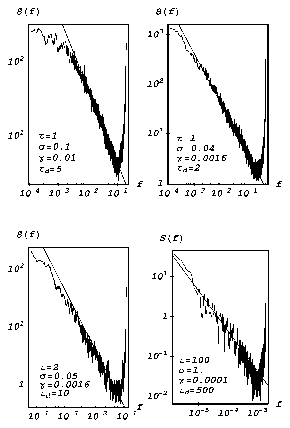}
  \end{center}
  \caption{Power spectral density versus frequency calculated from the
model described by Eqs. (3), (4), (7), (8). The main parameters
defining the model are: $\overline{\tau },\sigma ,\gamma $. Time
scale $\tau _{d}$ determines the highest frequency $f\leq
\frac{1}{\tau _{d}}$ under consideration. The sinuous curves
represent the results of numerical simulations averaged over five
realizations, and the straight lines represent the analytical
spectra described by Eqs. (5) and (6).}
  \label{Picture1}
\end{figure}

New discrete series are equivalent to the initial sequence of $\delta $
functions, when $\tau _{d}\gtrsim \overline{\tau }$ whit the same values of $%
\gamma $ and $\sigma $. Numerical results of $S(f_{s})$ calculated
with various parameters are shown in Fig.~\ref{Picture1}. Multiple
numerical calculations confirm full correspondence of the discrete
model (DM) defined here with earlier introduced by B.~Kaulakys and
T.~Me\v{s}kauskas \cite{9}.

The model can produce the $1/f$- like spectrum in an arbitrarily wide range
of frequencies $f_{1}<f<f_{2},f_{\tau }$ and is free from unphysical
divergence of the spectrum at $f\rightarrow 0$; for $f<f_{1}$:
\begin{equation}
S(f)=\overline{\tau }^{-2}\frac{2\sigma ^{2}/\overline{\tau }\gamma ^{2}}{%
1+\sigma ^{4}/4\overline{\tau }^{2}\gamma ^{4}}.
\end{equation}
Due to the long memory random process, defining transit time sequence $%
t_{k}, $ the model describes long time correlations quantified in power
spectral density $S(f)\backsim 1/f$. This model may be also generalized for
non-Gaussian distribution of the periods $\tau _{k}$. Then
\begin{equation}
S(f)=2\overline{\tau }^{-1}\psi (0)/f,
\end{equation}
where $\psi (\tau )$ is the distribution density of periods $\tau _{k}$.
This makes the model applicable to the wide variety of stochastic processes,
which have well defined distribution function $\psi (\tau )$ in the vicinity
of $\tau _{k}=0$. Numerical calculations confirm that, when $\psi (0)=0$,
the dependance of the power spectral density on frequency appears as $%
S(f)\backsim 1/f^{3/2}$ \cite{10}.

\begin{figure}[tbp]
  \begin{center}
  \includegraphics [width=1.0\hsize] {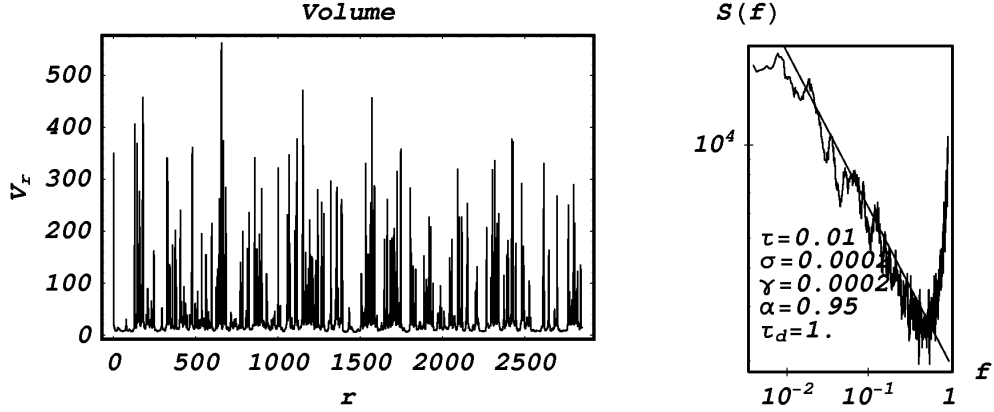}
  \end{center}
  \caption{Volume $V_{r}$ versus the number $r$ of time interval $\tau
_{d}$ and power spectral density $S(f)$ versus frequency $f$,
calculated
from the generalized discrete model (GDM) with $\sigma =0.0002,$ $\overline{%
\tau }=0.01,$ $\alpha =0.95,\gamma =0.0002,\tau _{d}=1.$  $S$ is
averaged over five realizations. The straight lines represent the
fractional power law $S(f)=2\ast 10^{3}/f^{1/2}$.}
  \label{Picture2}
\end{figure}

One more generalization of the model is needed for applications in financial
markets and other self-affine time series with a power-law dependence on
frequency $S(f)=f^{-\beta }$ with $\beta \simeq 1/2.$ We will strengthen
high frequencies and will account for positive playback of $\tau _{k}$
increment by adding the term $\alpha \Delta \tau _{k-1}=\alpha (\tau
_{k-1}-\tau _{k-2})$ to the $\tau _{k}$ recurrent expression:
\begin{equation}
\tau _{k}=\tau _{k-1}+\alpha (\tau _{k-1}-\tau _{k-2})-\gamma (\tau _{k-1}-%
\overline{\tau })+\sigma \varepsilon _{k}.
\end{equation}
Note that this new term changes autoregressive process AR(1) to
the higher one AR(2). Multiple numerical calculations with the
generalized discrete model (GDM) exhibit dependance of $\beta $ on
$\alpha $ and other parameters of the model: $\sigma
,\overline{\tau },\gamma $. We demonstrate an example of numerical
calculation with GDM in Fig.~\ref{Picture2}, which exhibits clear
fractional power law with $\beta =1/2$ of the power spectral
density.

\section{Application to the financial market}

\begin{figure}[tbp]
  \begin{center}
  \includegraphics[width=1.0\hsize]{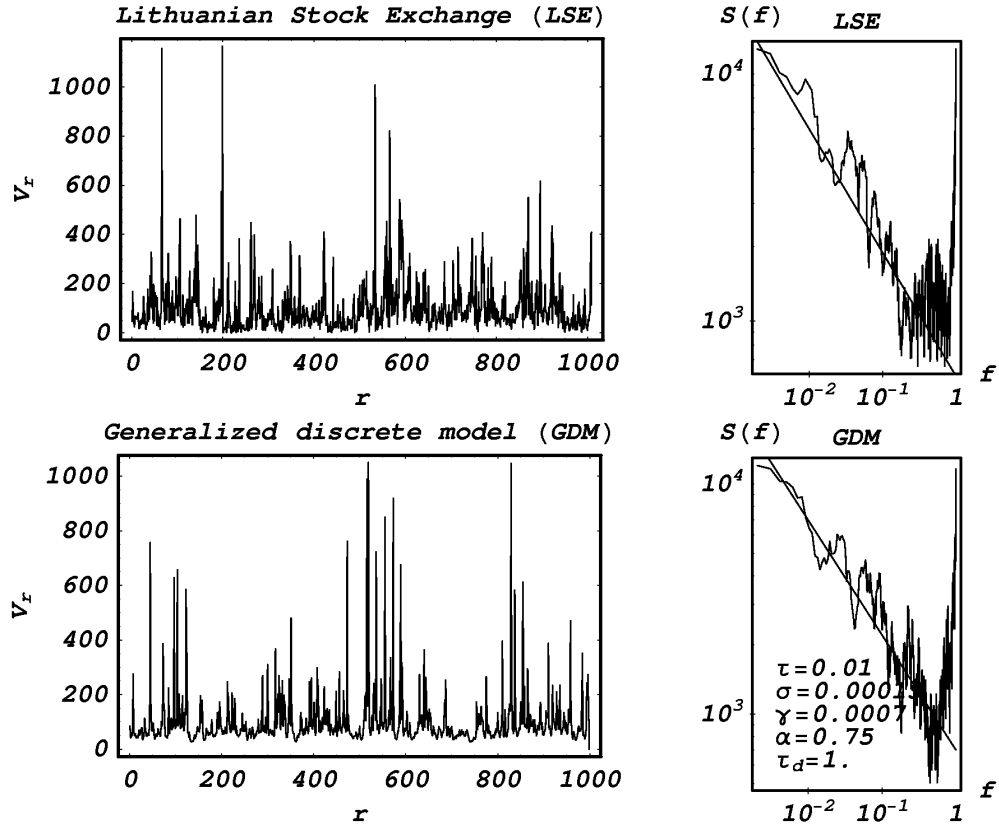}
  \end{center}
  \caption{\textbf{(LSE)} The volume of shares $V_{r}$ , included in
the index LITIN of Lithuanian Stock Exchange, normalized to an
average of 100 trades per $\tau _{d}=1day$ versus the number of
day traded and corresponding power spectral density $S(f)$ versus
frequency $f$ calculated from Fast Fourier Transform of discrete
data. \textbf{(GDM)} Volume $V_{r}$ versus the number $r$ of the
time interval $\tau _{d}=1$ and power spectral density $S(f)$
versus frequency $f$, calculated from the generalized
discrete model (GDM) with $\sigma =0.00015,$ $\overline{\tau }=0.01,$ $%
\alpha =0.75,\gamma =0.0007,\tau _{d}=1.$ The straight lines
approximating
the power spectral\ density curves represent the fractional power law $%
S(f)=700/f^{1/2}.$}
  \label{Picture3}
\end{figure}

An important quantity that characterizes the dynamics of price
movement in financial markets is the number of shares $V_{r}$
(share volume) traded in a time interval $r\tau _{d}<t<(r+1)\tau
_{d}$. The statement ``It takes volume to move stock prices''
accumulates very general idea that statistical properties of
financial markets are enclosed in time series of share volume.
Very direct confirmation of this statement and quantitative
investigation of the largest 1000 stocks in three major US stock
markets was recently presented in \cite{6}. This work provides an
evidence that long range correlations in share volume and price
volatility are largely due to those of the number of trades
$N_{r}$ in time interval $\tau _{d}$. Close correlation between
$N_{r}$ and $V_{r}$ is imposed by the relation:
\begin{equation}
V_{r}\equiv \sum_{i=1}^{N_{r}}q_{i}
\end{equation}
and very weak correlation in time sequence of share volume per transaction $%
q_{i}$. These results suggest us to apply GDM as a model for the time series
of share volume traded in financial markets , with a simple assumption that
in the first approach the average $\left\langle q_{i}\right\rangle $ can be
included into the normalization factor. This mean that simple relation $%
V_{r}=N_{r}\left\langle q_{i}\right\rangle $ enables us to make
comparison of GDM to the variety of real market data. In
Fig.~\ref{Picture3}. we demonstrate comparison of Lithuanian Stock
Market data with numerical results from GDM. The volume of shares
included in the index LITIN is normalized to average of 100 trades
per $\tau _{d}=1\mathrm{day}$ . Note that despite the model
simplicity it serves as market data generator and reproduces the
main statistical property of the system, i.e., the power law
dependence ($\beta =1/2$) on the frequency of the power spectral
density.

\section{Conclusion}

The empirical evidence provided by Gopikrishnan \emph{et al} \cite{6}, that
the number of transactions $N_{r}$ in the subsequent time intervals $\tau
_{d}$ define long range correlations of share volume traded, enabled us to
apply simple model of a $1/f$ noise \cite{9}\ and to reproduce long-range
correlations of the share volume traded in financial markets. We generalized
the KM model by integrating the sequence of pulses in a conventional time
interval $\tau _{d}$ and replacing the recurrent time $\tau _{k}$ stochastic
process from AR(1) to the AR(2) one. Numerical calculations with the
generalized discrete model (GDM) reproduce power spectral density $S(f)$
scaled as power of frequency $1/f^{\beta }$ for various values of $\beta $,
including $\beta =1/2$ for applications in financial markets. Further
investigation of the model with its possible applications in financial time
series is in progress.

\section*{Acknowledgment}

The author gratefully acknowledges numerous and long discussions with
Bronislovas Kaulakys regarding inherent origin of $1/f$ noise and
applicability of KM model.


\begin{thebibliography}{99}
\bibitem{1}  \textbf{J. D. Farmer,} Computing in Science \& Engineering, V.
1, p. 26 (1999).

\bibitem{2}  \textbf{R. N. Mantegna and H. E. Stanley, }\textit{Introduction
to Econophysics: Correlations and Complexity in Finance}, (UK,Cambridge,
Cambridge Univ. Press., 1999).

\bibitem{3}  P. \textbf{Gopikrishnan et al. }Phys. Rev. E, V. 60, p. 5305
(1999).

\bibitem{4}  \textbf{V. Plerou et al., }Phys. Rev. E,\textbf{\ }V. 60, p.
6519 (1999)

\bibitem{5}  \textbf{Y. Liu et al., }Phys. Rev. E, V. 60, p. 1390 (1999).

\bibitem{6}  P. \textbf{Gopikrishnan et al., }Phys. Rev. E, V. 62 (4), p.
R4493-R4496 (2000).

\bibitem{7}  \textbf{B. B. Mandelbrot,} Science, V. 156, p. 636-638 (1967).

\bibitem{8}  \textbf{B. B. Mandelbrot, J. W. van Ness, }SIAM Rev., V. 10, p.
422-437 (1968).

\bibitem{9}  \textbf{B. Kaulakys and T. Me\v{s}kauskas, }Phys. Rev. E, V 58,
p.7013 (1998).

\bibitem{10}  \textbf{B. Kaulakys and T. Me\v{s}kauskas, }Nonlinear
Analysis: Modelling and Control, Vilnius, IMI, 1999, No 4
\end{thebibliography}
\end{document}